\documentclass[twocolumn]{aastex62}
\usepackage{natbib}
\graphicspath{{./}{figures/}}
\usepackage{color,subfigure,lineno} 

\shorttitle{VODKA: Off-nucleus AGN at low-z}
\shortauthors{Shen et al.}

\begin{document}

\title{Varstrometry for Off-nucleus and Dual sub-Kpc AGN (VODKA): How Well-centered Are Low-$z$ AGN?}


\author[0000-0003-1659-7035]{Yue Shen}
\altaffiliation{Alfred P. Sloan Research Fellow}
\affiliation{Department of Astronomy, University of Illinois at Urbana-Champaign, Urbana, IL 61801, USA}
\affiliation{National Center for Supercomputing Applications, University of Illinois at Urbana-Champaign, Urbana, IL 61801, USA}

\author[0000-0003-4250-4437]{Hsiang-Chih Hwang}
\affiliation{Department of Physics and Astronomy, Johns Hopkins University}

\author{Nadia Zakamska}
\affiliation{Department of Physics and Astronomy, Johns Hopkins University}

\author[0000-0003-0049-5210]{Xin Liu}
\affiliation{Department of Astronomy, University of Illinois at Urbana-Champaign, Urbana, IL 61801, USA}
\affiliation{National Center for Supercomputing Applications, University of Illinois at Urbana-Champaign, Urbana, IL 61801, USA}

\begin{abstract}
Off-nucleus active galactic nuclei (AGN) can be signposts of inspiraling supermassive black holes (SMBHs) on galactic scales, or accreting SMBHs recoiling after the coalescence of a SMBH binary or slingshot from three-body interactions. Because of the stochastic variability of AGN, the measured photocenter of an unresolved AGN-host system will display astrometric jitter that depends on the off-nucleus distance of the AGN, the total photometric variability of the system, and the AGN-host contrast. Here we use the precision astrometry from Gaia DR2 to constrain the off-nucleus population of a low-redshift ($0.3<z<0.8$) sample of unobscured broad-line AGN drawn from the SDSS with significant host contribution and photometric variability. We find that Gaia DR2 already provides strong constraints on the projected off-nucleus distance in the sub-kpc regime at these redshifts: $99\%$, $90\%$ and $40\%$ of AGN must be well-centered to  $<1$ kpc, $<500$ pc and $<100$ pc, respectively. Limiting the sample to the most variable subset constrains $>99\%$ of AGN to be well-centered below 500 parsec. These results suggest that genuine off-nucleus AGN (offset by $>$ a few hundred pc) must be rare at low redshift. Future Gaia releases of time series of photocenter and flux measurements, improved treatments for extended sources and longer baselines will further tighten these constraints, and enable a systematic full-sky search for rare off-nucleus AGN on $\sim 10-1000$ pc scales.   
\end{abstract}


\section{Introduction}\label{sec:introduction}

The formation and evolution of supermassive black holes at the center of galaxies are still not fully understood. Galaxy mergers are often invoked as an important route to fuel SMBHs and to produce pairs of SMBHs at different separations from tens of kpc scales to the parsec regime where the two SMBHs become gravitationally bound. An AGN may be observed as an off-nucleus source during the inspiraling phase of the two merging galaxies \citep[e.g.,][]{barth08,Comerford2014}. If the SMBH binary eventually coalesces, the merger remnant may receive a kick from anisotropic gravitational wave radiation \citep{Baker2006,campanelli07}. If the recoiled SMBH accretes from the ambient gas, it may also be observed as an off-nucleus AGN \citep[e.g.,][]{loeb07,Komossa2012}. Therefore the observed statistics of off-nucleus AGN provide important constraints on AGN fueling processes in galaxy mergers as well as the population of recoiling SMBHs following binary coalescence \citep[e.g.,][]{Blecha2016,Raffai2016,Tremmel2018}. 

However, it is extremely difficult to observe this population of off-nucleus SMBHs, given the stringent spatial resolution requirement. If AGN are more likely triggered in late stages of mergers then their offset from the galactic center is expected to be below $\sim 1$ kpc. Likewise, given typical recoil velocities (hundreds of ${\rm km\,s^{-1}}$) of SMBH coalescence \citep{bogdanovic07,Miller2013}, the kicked SMBH cannot travel far from the galactic center \citep[e.g.,][]{merritt04,Komossa2008,Blecha2011}. There are additional, observational difficulties in observing off-nucleus AGN. For example, if the AGN is optically unobscured and contributes significantly to the total flux of the unresolved AGN$+$host system, it may be challenging to determine the host center properly in the presence of a bright point source. 

\citet{Comerford2015} used the combination of X-ray detection and optical/IR imaging to detect off-nucleus AGN and found several candidates on $\gtrsim$ kpc-scales \citep[also see][]{Barrows_etal_2016}. Other serendipitously discovered kpc-scale off-nucleus AGN candidates \citep[e.g.,][]{Civano_etal_2010,Comerford_etal_2009b,Kim_etal_2017} also relied on the high imaging resolution from HST. The demand for multi-wavelength coverage and/or the limited positional accuracy of most imaging facilities ($\gtrsim$0\farcs1-0\farcs5) prevents a systematic search of off-nucleus AGN in the sub-kpc regime. While candidate off-nucleus AGN can also be identified with kinematic offset in the velocity of the broad or narrow emission lines \citep[e.g.,][]{loeb07,Liu2014,Comerford2014,Runnoe_etal_2017}, the interpretation is not straightforward given the complexity of the emission-line region kinematics, and usually these kinematically-identified candidates have no constraint on the spatial offset. For these reasons, there is currently no statistical constraint on the off-nucleus AGN population on scales between $\sim 10$ pc and $\sim 1$ kpc.


In this work, we present a different approach to constrain the population of off-nucleus AGN in a systematic fashion, and in particular to cover the sub-kpc regime. The working principle was described in \citet[][hereafter Paper I]{Hwang_etal_2019}: consider an unresolved AGN$+$host system (i.e., the separation $D$ is smaller than the full-width-at-half-maximum of the point spread function PSF), since essentially all AGN vary stochastically, an off-nucleus AGN produces variability-induced astrometric jitter in the photocenter of the unresolved system. This technique was dubbed {\tt varstrometry} in Paper I for ``intrinsic variability induced astrometric jitter''. The same idea was applied to identify variable binary stars \citep[e.g.,][]{Wielen_1996,Pourbaix_etal_2003}, in which the method was dubbed `variability-induced movers' (VIMs). First extragalactic applications of this technique include \citet{Shen2012a} on constraining the sizes of the broad-line region and torus in quasars, and \citet{Liuyuan2015} on identifying dual AGN.

The expected astrometric signal from sub-kpc off-nucleus AGN falls well within the reach of the precision astrometry enabled by the Gaia mission \citep{Gaia2016}. For $\sim$kpc and smaller scales and in the redshift range considered here ($0.3<z<0.8$, see \S\ref{sec:data}), Gaia does not resolve the AGN and the host into separate sources and measures a single-source photocenter, thus validating the use of {\tt varstrometry}. More importantly, Gaia is an all-sky optical survey to $G\lesssim 21$ mag, which then allows a systematic exploration of off-nucleus AGN with unprecedented statistics and spatial constraints. 

The paper is organized as follows. In \S\ref{sec:data} we describe the methodology and the data we use for this study. We present our main results in \S\ref{sec:gaia} and discuss their implications in \S\ref{sec:disc}. We summarize our findings and discuss future prospects in \S\ref{sec:con}. Throughout this paper we focus on optically unobscured, broad-line AGN, for which we can apply this {\tt varstrometry} technique, and all physical separations are the projected separation. A flat $\Lambda$CDM cosmology is adopted throughout with $\Omega_\Lambda=0.7$ ($\Omega_0$=0.3) and $H_0=70\,{\rm km\,s^{-1}Mpc^{-1}}$. 

\section{Methodology and Data}\label{sec:data}

In Paper I we provided a detailed description of the working principles of {\tt varstrometry}. In brief, for a single AGN that is offset from the center of the host galaxy at a distance $D$, the stochastic photometric variability of the AGN introduces an astrometric shift of the measured photocenter of the unresolved system. This astrometric shift is aperiodic, along a bound line segment, and correlates with the photometric variability. If we have the full time series of the photocenter and flux measurements (both measured in the same bandpass), and a one-time estimate of the AGN-host contrast at any epoch, the off-nucleus distance $D$ can be derived from a regression fit between the photometric and astrometric time series (Paper I) that utilizes all the data points. However, even if we only have estimates of the root-mean-squares (RMS) of the photocenter and flux variations, we can still derive the off-nucleus distance $D$ from the two RMS quantities (cf. Paper I).

The photocenter of the unresolved system is the first moment of the AGN+host image convolved with the PSF. If we assume that the PSF is axisymmetric and stable (spatially and temporally), the photocenter is simply the ``center of light'', or the flux-weighted average of the individual centroids of the AGN and the host. This conclusion is valid regardless of the extended nature of the host, as long as the aperture used to compute the centroid encloses most of the flux. This was also demonstrated with simulated images in Paper I. Taylor expanding the photocenter with respect to flux variation, we derive the following RMS relation (same as eqn.~ 5 in Paper I):

\begin{equation}\label{eqn:key1}
    \sigma_{\rm astro}=D\frac{q}{1+q}\frac{\sqrt{\langle\Delta f^2\rangle}}{\bar{f}}\ ,
\end{equation}
where $q\equiv f_{\rm host}/\bar{f}_{\rm AGN}$ is the mean host-to-AGN contrast, $\sigma_{\rm astro}$ is the RMS in photocenter and $\sqrt{\langle\Delta f^2\rangle}/\bar{f}$ is the fractional photometric RMS of the system. Only the variability of the AGN contributes to $\langle\Delta f^2\rangle$, but both the AGN and the host contribute to the total mean flux ${\bar{f}}$. This equation is derived to the leading order in $\Delta f$.  

Eqn.~(\ref{eqn:key1}) is derived for an idealized PSF. If the PSF is asymmetric but stable, the overall photocenter will have a constant offset, which does not affect the RMS relation (\ref{eqn:key1}). If the PSF varies spatially or temporally, and if the aperture used to compute the photocenter misses significant flux due to the extended host morphology, there may be additional systematic errors on the photocenter RMS. However, as discussed below, systematic errors are included in the photocenter RMS estimate provided by Gaia, and therefore we have even more stringent constraints on the pair separation following Eqn.~(\ref{eqn:key1}).

Normally $\sigma_{\rm astro}$ and $\sqrt{\langle\Delta f^2\rangle}/\bar{f}$ should be directly computed from the time series of photocenter (used by Gaia to derive astrometric solutions) and light curves, which are currently unavailable for Gaia DR2. Instead, we use quantities that are released in Gaia DR2 as surrogates for the astrometric and photometric RMS. 

We estimate the photometric RMS using the reported Gaia mean flux uncertainties and the number of photometric observations, i.e.,
\begin{equation}
    \sigma_G={\tt phot\_g\_mean\_flux\_error}\sqrt{{\tt phot\_g\_n\_obs}}\ ,
\end{equation}
which has been demonstrated in Paper I to be a good proxy for the actual photometric RMS. In calculating the intrinsic photometric RMS for the system (AGN+host), we subtract in quadrature the RMS from stars with the same mean $G$-band flux, which represents the measurement (statistical plus instrumental) uncertainties. Thus we derive the total intrinsic photometric RMS variability for the system. The caveat is that the photometric uncertainty floor derived from stars is only well determined for objects brighter than $G=20$, which limits the magnitude range for our usable AGN sample.  

We estimate the astrometric RMS $\sigma_{\rm astro}$ using the quantity {\tt astrometric\_excess\_noise} (in units of mas). As discussed in \citet{Lindegren_etal_2012}, this quantity describes the extra noise term added to the statistical uncertainties of the photocenter positions, and therefore represents the intrinsic astrometric RMS (astrophysical or systematic). Potential contributions to this excess astrometric noise could come from off-nucleus AGN, variability in sub-kpc optical jets \citep[e.g.,][]{Petrov_etal_2019}, or systematics in the data processing and assumptions, for example, unmodeled PSF variations. In particular, in Paper I we show that extended morphology of the target (e.g., host galaxies) in some cases can lead to large astrometric excess noise. This is because the scanning direction of Gaia is different during different passes and an extended low-redshift galaxy may have significant flux outside the nominal window used by Gaia to measure the photocenter. This complication introduces additional RMS error in the photocenter measurement. This is a limitation of Gaia DR2, and will be improved in future Gaia releases that will have better treatment of extended sources. For these reasons, the quantity {\tt astrometric\_excess\_noise} from Gaia DR2 should be treated as an upper limit for the astrophysical astrometric jitter due to {\tt varstrometry} in off-nucleus AGN. 

Equation (\ref{eqn:key1}) implies that an appropriate host-AGN contrast will enhance the {\tt varstrometric} signal. If $q$ is too small, the AGN dominates the light and the photocenter shifts are small. On the other hand, if the galaxy dominates the light, the photometric variability is diluted and the expected {\tt varstrometric} signal is small as well. The optimal situation is when the host-AGN contrast is moderate and the AGN variability is substantial. 

To maximize the constraining power from Gaia data, we choose a low-redshift AGN sample from SDSS \citep{Sun_Shen_2015}. These objects are spectroscopically confirmed as unobscured, broad-line AGN from optical spectroscopy. \citet{Sun_Shen_2015} performed spectral decomposition using a Principle Component Analysis approach \citep[e.g.,][]{Vandenberk_etal_2006} on these objects and estimated the host fraction in these low-redshift quasars. Compared to high-luminosity and high-redshift quasars, this sample contains AGN with significant host contribution covered by the Gaia $G$ band, and significant photometric variability to be measured by Gaia, and therefore is an ideal sample for our study. In total there are 32,040 unique AGN in the sample, among which 29,769 have been successfully matched to Gaia to within 0.3\arcsec\ ($\sim 1$ kpc) from the SDSS position with a single Gaia detection. 

There are 263 AGN with two Gaia sources within 3\arcsec, including one AGN with two Gaia sources within 0\farcs3, and 2 AGN with three Gaia sources within 3\arcsec. These are potentially dual AGN on kpc separations or AGN$+$star pairs, and are further discussed in \S\ref{sec:disc2}. 

There are also 84 AGN with a single Gaia match at $>$0\farcs3, or $\sim 0.3\%$ of all the singly-matched AGN. Considering the typical SDSS positional accuracy of $\sim$ 0\farcs1, this fraction is consistent with that of 3$\sigma$ astrometric offset expected from statistical uncertainties. Indeed, there are no significant differences in the Gaia photometric color and parallax/proper motion distributions between these 84 AGN and the bulk of the singly-matched AGN. For simplicity we remove these 84 objects from futher consideration. 

Our parent sample is therefore the 29,769 unique Gaia sources that enclose the central $\sim 1$ kpc region of the SDSS AGN. For these objects, the AGN$+$host is unresolved in Gaia DR2 (i.e., Gaia measures a single source photocenter), and thus ideal for probing the sub-kpc regime of off-nucleus AGN with {\tt varstrometry}. The high completeness of Gaia detection (29,769/32,040) of low-redshift SDSS AGN also ensures that our statistical constraints are not biased against potential off-nucleus AGN. Indeed most, if not all, of these non-detections are simply due to the faintness of the object. 

To obtain a clean sample of AGN to constrain the off-nucleus AGN population, we impose the following additional criteria:
\begin{enumerate}
\item[$\bullet$] We limit the host-to-total fraction at rest-frame 5100\,\AA\ measured in \citet{Sun_Shen_2015} to be 10-90\% to ensure appropriate contrast ratios between the AGN and the host (this constraint cuts the sample down to 21,839 objects);
\item[$\bullet$] We limit the sample to $G<20$ so that we have reasonably well-measured proxy for the photometric RMS from Gaia DR2 (down to 18,709 objects); 
\item[$\bullet$] We restrict the sample to $0.3<z<0.8$ to eliminate very low-redshift objects with potentially overestimated {\tt astrometric\_excess\_noise} due to the extended morphology of the host galaxy (down to 9,872 objects); 
\item[$\bullet$] We require a minimum photometric RMS of 5\% to ensure the variability estimates are reliable and to enhance the constraints from {\tt varstrometry} (down to 8,210 objects).  
\end{enumerate}

Importantly, none of these additional cuts introduces selection biases for or against the sub-kpc off-nucleus AGN population that we can think of. Our final cleaned sample contains 8,210 AGN, which is large enough to constrain the distribution of off-nucleus distance.  

We assume that all objects in this sample are single AGN, since they dominate by number the much rarer dual AGN population \citep[e.g.,][]{Liu_etal_2011}. However, if the optical light of the unresolved system is dominated by two AGN, the expected astrometric RMS has a similar dependence on $D$ and total photometric RMS (cf. eqn.\ 3 in Paper I). Therefore the inclusion of a small fraction of potential dual AGN in our sample does not affect our results.   

\begin{figure*}
 \includegraphics[width=0.48\textwidth]{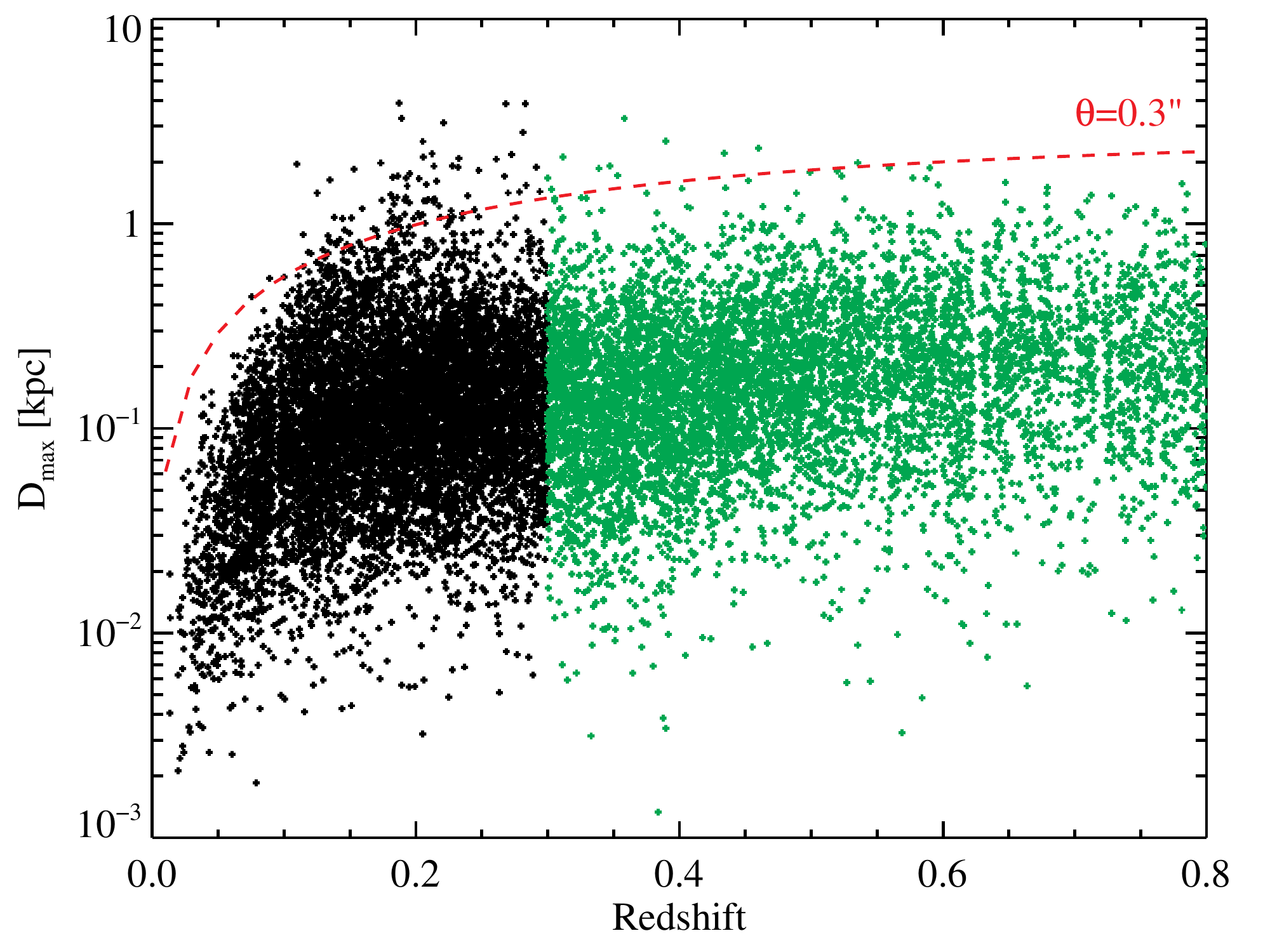}
 \includegraphics[width=0.48\textwidth]{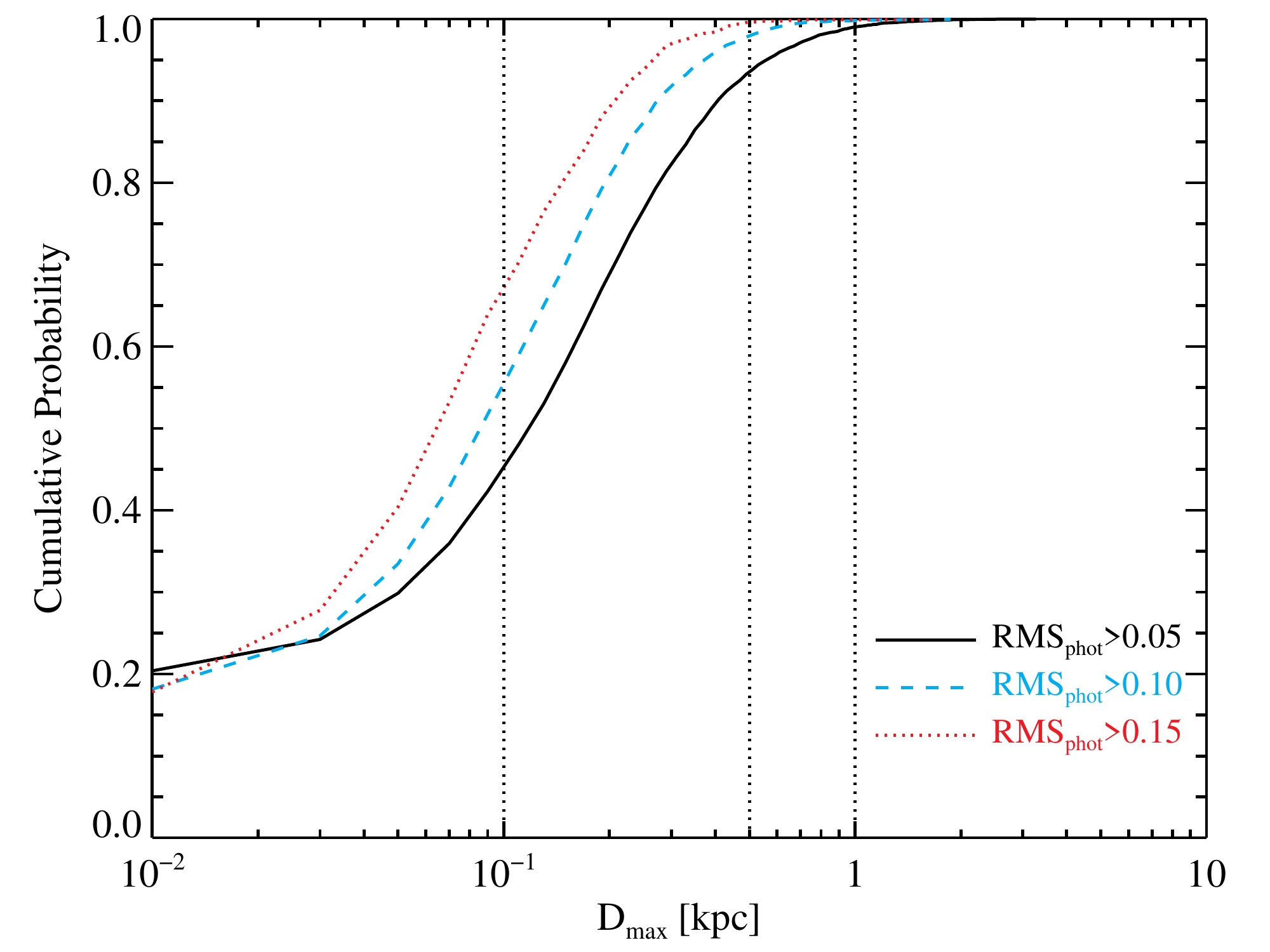}
 \caption{{\it Left}: individual upper limits on the off-nucleus distance of AGN as a function of redshift. The green points are for the clean sample ($0.3<z<0.8$) for which we derive statistical constraints. The red dashed line corresponds to the projected physical length at 0\farcs3. There are not many objects scattered beyond the red line, which would have been resolved into multiple Gaia sources. The excess of objects beyond the red line at $z<0.3$ is likely caused by the imperfect treatment of extended sources in Gaia DR2. {\it Right}: cumulative probability distributions of the upper limit on the off-nucleus distance for the clean AGN sample (black line) and two subsets of the clean AGN sample with more stringent thresholds on the photometric variability (cyan and red lines). \label{fig:dmax}}
 \end{figure*}

\section{Results}\label{sec:gaia}

We use {\tt astrometric\_excess\_noise} as a strict upper limit on the astrometric RMS due to off-nucleus AGN. Figure \ref{fig:dmax} (left) displays the upper limit on the off-nucleus distance, $D_{\rm max}$, estimated from Eqn.\ (\ref{eqn:key1}) for individual objects as a function of redshift. Our cleaned sample is indicated by the green points. Below $z=0.3$ there is an excess of objects with $D_{\rm max}>1$ kpc compared with the $z>0.3$ subset: these are due to large {\tt astrometric\_excess\_noise} potentially caused by the extended source morphology instead of intrinsic astrometric RMS (Paper I). The red curve indicates the projected physical separation of 0\farcs3. As a sanity check, we should not expect many objects scattered beyond the red line, since otherwise these objects will likely have multiple detections in Gaia DR2, contrary to the selection criteria of our sample. However, it is possible that a small number of AGN have $>1$ kpc separation from the host center but are still unresolved in Gaia DR2 due to scanning strategies and processing details.  

Figure \ref{fig:dmax} (right) displays the cumulative probability distribution of the upper limits on the off-nucleus distance for our AGN sample (black line). $99\%$, $90\%$ and $40\%$ of these AGN must be well-centered to $<1$ kpc, $<500$ pc and $<100$ pc, respectively. These are by far the strongest (and the first) statistical constraints on the off-nucleus AGN population in the sub-kpc regime and for these redshifts.  

\begin{figure}
 \includegraphics[width=0.48\textwidth]{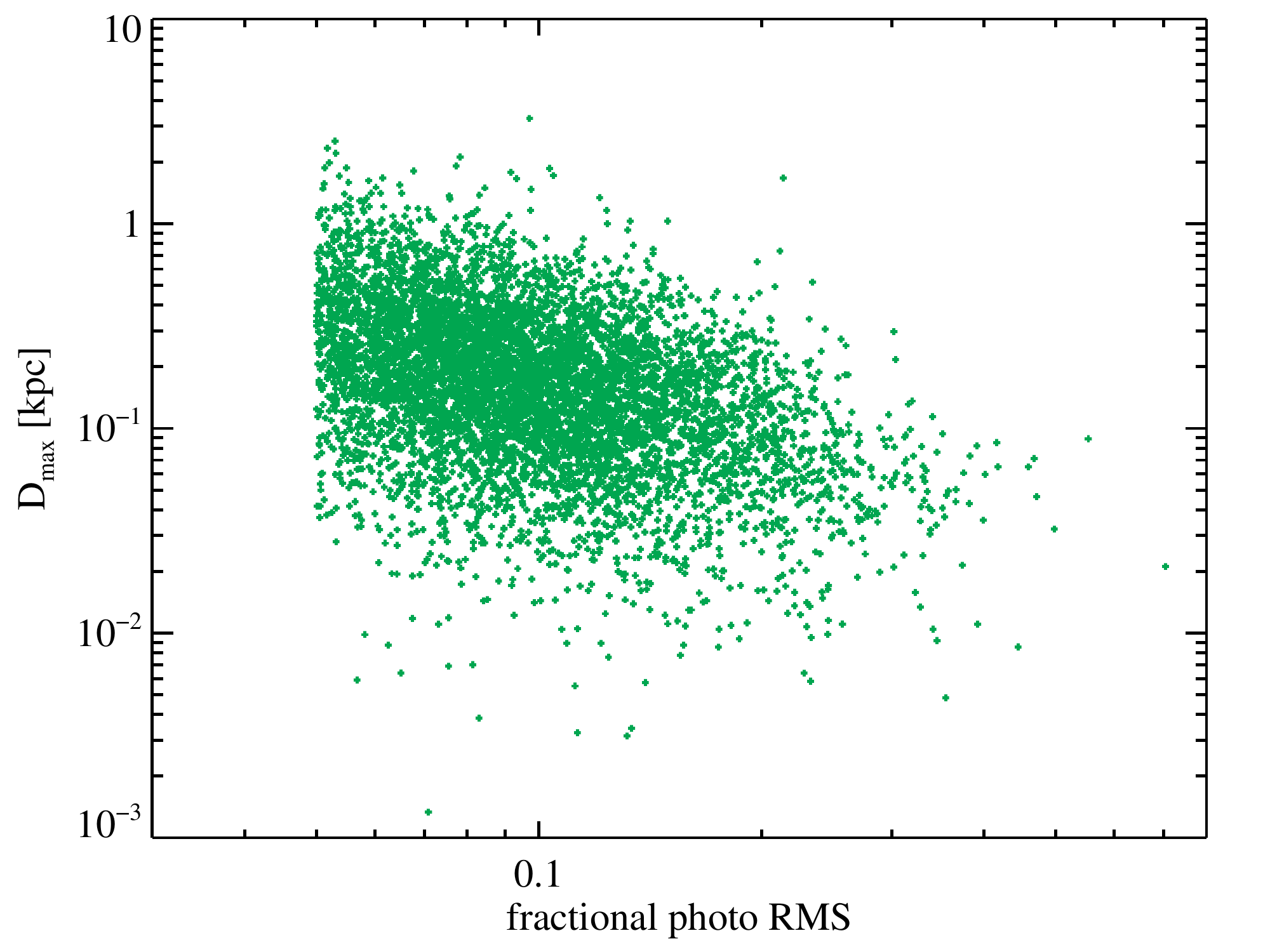}
 \caption{The relation between the upper limit on the off-nucleus distance and the fractional photometric variability for the clean AGN sample. More variable AGN have tighter constraints on the upper limit. \label{fig:dmax2}}
 \end{figure}

Since we do not expect there is any significant correlation between the photometric variability and off-nucleus distance, we can impose more stringent thresholds on the minimum photometric variability to tighten the constraints on off-nucleus distance (cf., Eqn.\ \ref{eqn:key1}). Fig.\ \ref{fig:dmax2} displays the upper limits we derived for individual AGN in our sample as a function of the fractional photometric RMS variability. Indeed the upper limits on $D$ become tighter for more variable systems. In Fig.\ \ref{fig:dmax} (right) we show the cumulative probability distributions of $D_{\rm max}$ for two subsamples with larger fractional photometric RMS thresholds ($>10\%$ and $>15\%$), both of which still retain sufficient number statistics ($N>1000$). With these more variable subsamples of AGN, we were able to further limit the prospect of a substantial off-nucleus AGN population. For example, the most variable AGN subsample constrains $99.6\%$ of the population to be well-centered below 500 parsec. 

Finally, we confirm that there is no correlation between the astrometric RMS (converted to projected physical lengths) and the fractional photometric RMS. In cases where the off-nucleus AGN population has a characteristic separation (e.g., $\sim 500$ parsec) from the host center, we would expect a positive correlation between the astrometric and photometric RMS (cf. Eqn.\ \ref{eqn:key1}). The lack of such a correlation is indication that these AGN are well-centered or there is no preferred off-nucleus distance. 



\section{Discussion}\label{sec:disc}

\begin{figure*}
 \includegraphics[width=0.48\textwidth]{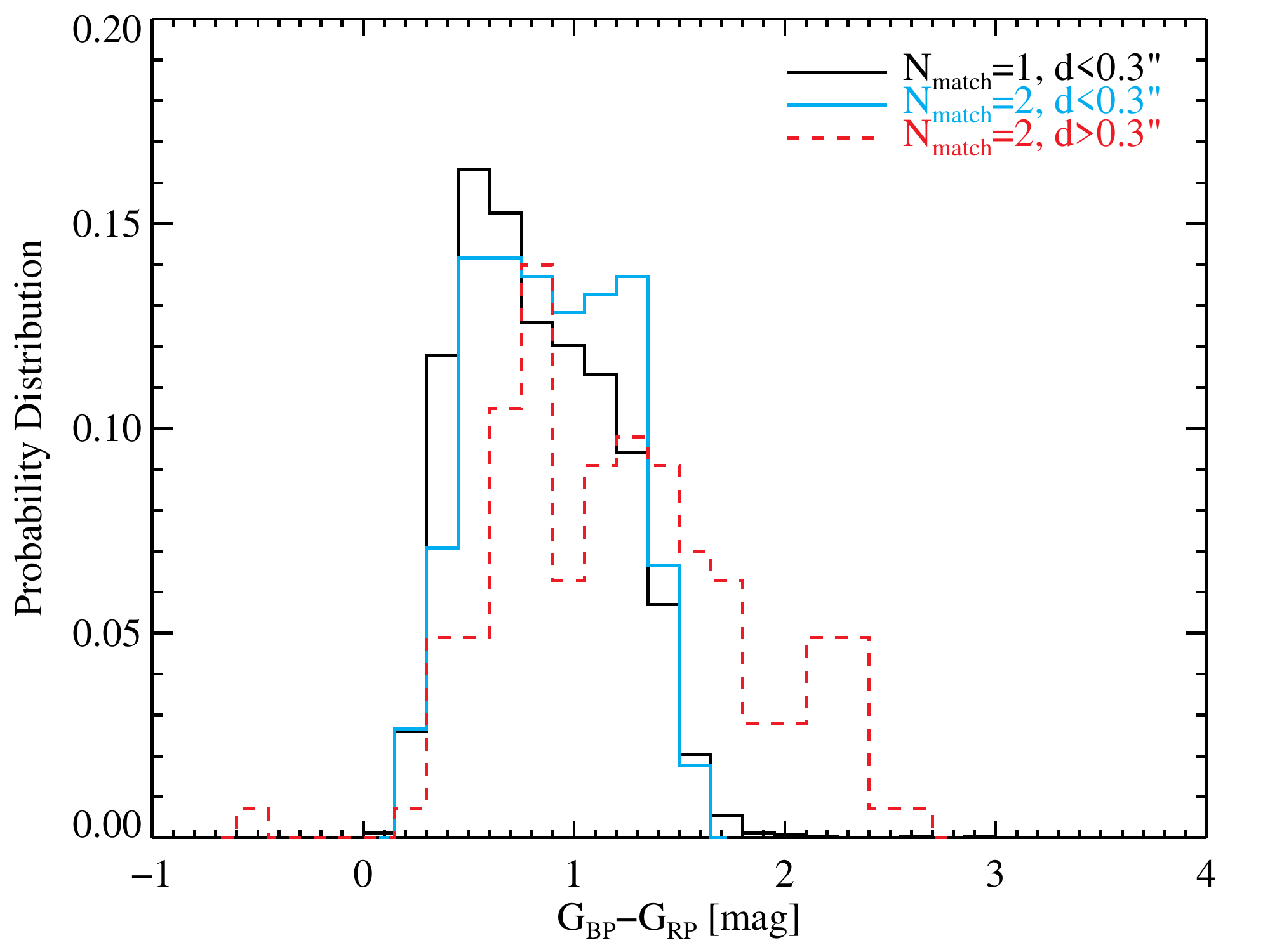}
 \includegraphics[width=0.48\textwidth]{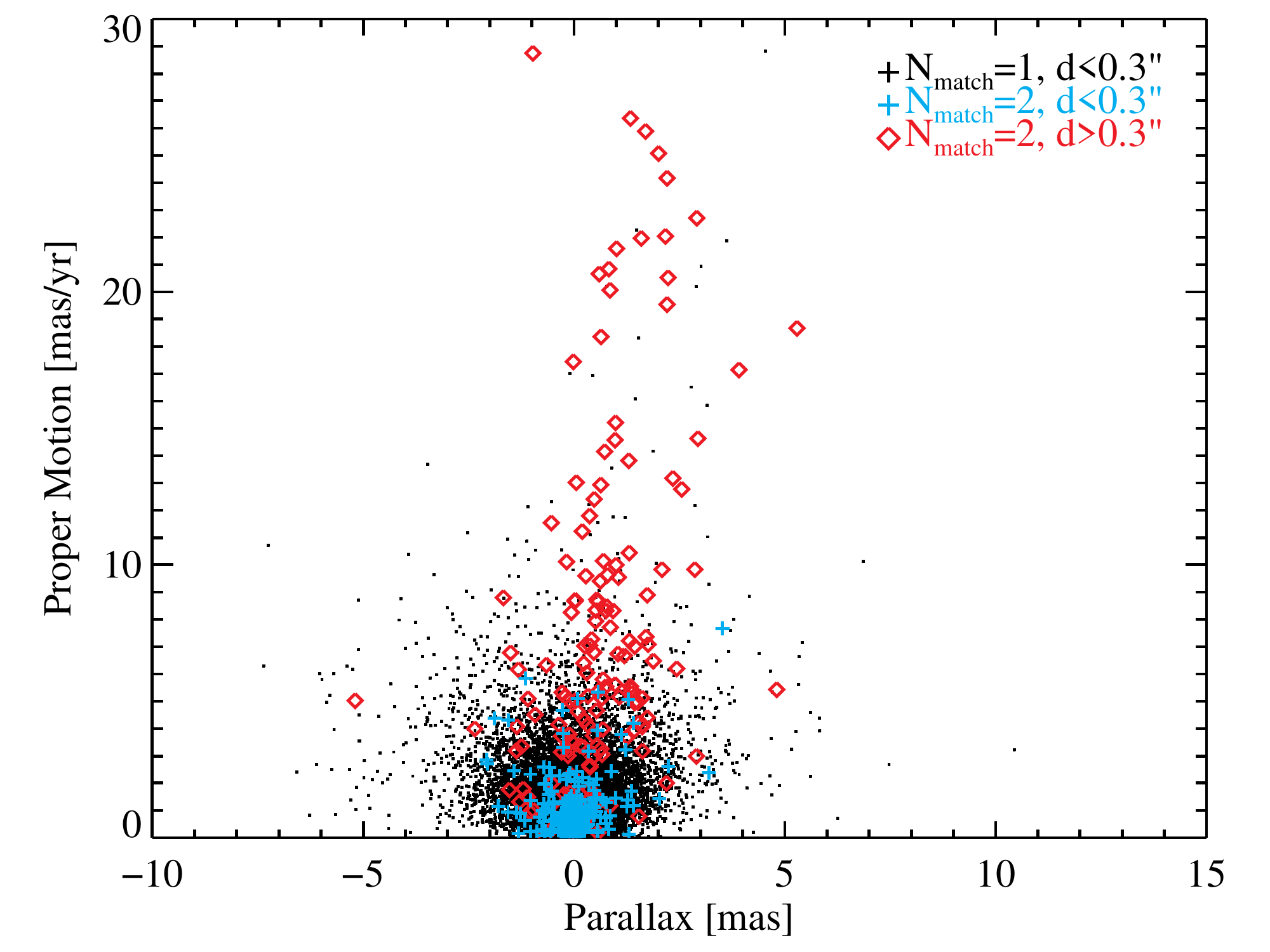}
 \caption{{\it Left}: probability distributions of the Gaia $G_{\rm BP}-G_{\rm RP}$ color for the parent AGN sample with single Gaia matches within 0\farcs3 (black), Gaia sources that are within 0\farcs3 of the SDSS position in multiply-matched cases (cyan) and those that are more than 0\farcs3 away from the SDSS position (red). {\it Right}: measured parallaxes and proper motions for different samples, with the same color scheme as in the left panel. In multiply-matched cases, the Gaia source closest to the SDSS position corresponds to the AGN, while the more distant Gaia source is most likely a foreground star given its different distributions in color, proper motion and parallax.  \label{fig:multi_gaia}}
 \end{figure*}

\subsection{Off-nucleus AGN are rare at low redshift}

Because {\tt astrometric\_excess\_noise} is an upper limit on the intrinsic {\tt varstrometric} signal, we can only place upper limits on the off-nucleus distance $D$. Fortunately, the superb astrometric precision (mas to tens of mas) of Gaia enables stringent upper limits on $D$. It appears that the vast majority ($>90\%$) of the AGN in our sample are well-centered to within several hundred parsec. For about half of the sample we can further constrain the off-nucleus distances to less than 100 parsec. These constraints become much tighter when we restrict our sample to the most variable subset, assuming AGN variability is not correlated with off-nucleus distance. 

There are multiple implications of these results. First, we can rule out the existence of a large population of unobscured off-nucleus AGN at $z\lesssim 1$ that are more than a few hundred parsec away from the galactic center. Even if we assume all the objects in our sample with multiple Gaia source matches within 3" are off-nucleus or dual AGN (but see \S\ref{sec:disc2}), they only amount to less than 1\% of the sample. Either these off-nucleus SMBHs do not exist, or they are accreting at levels too low to be detectable. 

Second, if most low-redshift AGN were merger-induced, then our measurements suggest that we must be witnessing them at late stages below kpc-scales where the SMBH has already well aligned with the galactic center. Furthermore, if most of these mergers result in the coalescence of a binary SMBH and a gravitational recoil, the recoiling BH cannot travel beyond a few hundred parsec in most cases. Alternatively, our measurements suggest that most of these low-redshift AGN are not in mergers (either before or after the coalescence of the binary SMBH).

Another possibility is that most sub-kpc off-nucleus AGN are obscured and missed from our sample entirely. The abundances of unobscured and obscured AGN are comparable at these redshifts \citep[e.g.,][]{Reyes_etal_2008}, which is consistent with pure orientation effects of the Type 1/Type 2 dichotomy and suggests that there should be no significant difference in the off-nucleus AGN population between obscured and unobscured AGN. However, potential evolutionary effects in mergers may cause a preference of off-nucleus AGN being obscured.

\subsection{Targets of potential interest}\label{sec:disc2}

We now investigate cases where the SDSS AGN has multiple Gaia matches within 3\arcsec\ from the SDSS position. There are 263 AGN with two Gaia matches and 2 AGN with three Gaia matches.

Figure \ref{fig:multi_gaia} displays the distributions in Gaia color, parallax and proper motion, for the singly-matched objects (dominant in number) and the multiply-matched objects. For the latter case we also divide the matches at a separation of 0\farcs3. The population of Gaia sources with separations $<$0\farcs3 in the multiply-matched cases is similar to the singly-matched objects, suggesting they are genuine Gaia matches of the AGN. In contrast, the population of Gaia sources with separations $>$0\farcs3, which correspond to $>$kpc separations, are redder and have more significant proper motion (or parallax) measurements, suggesting they are mostly Milky Way stars in superposition. Thus the fraction of missed off-nucleus AGN on $>$kpc scales from our statistical analysis in \S\ref{sec:gaia} is negligible. To identify rare genuine off-nucleus AGN (or dual AGN) among these $\sim 250$ off-center Gaia sources would require additional resources and a case-by-case study, and thus is beyond the scope of this paper. 

For each of the two AGN with 3 Gaia matches within 3\arcsec, the brightest Gaia source is associated with the AGN with negligible offset from the SDSS position. The two additional Gaia sources in each case are fainter than the central source, and do not have measurements on Gaia color, proper motion or parallax. A detailed look at their optical images from PanSTARRS \citep{Chambers2016} suggests that in one system (SDSS J145706.93$+$494011.0 at $z=0.013$) the fainter additional Gaia sources are associated with knots in a ring structure around the central nucleus; the other system (SDSS J121135.93$+$354417.6 at $z=0.06$) may be a real triple AGN system, or more likely, Gaia-resolved clumps in its galactic structure.  

There is only one AGN (SDSS J133039.82$-$001035.7 at $z=0.238$), with two Gaia matches within 0\farcs3 from the SDSS position. The two Gaia sources are separated by 0\farcs5 and have no parallax/proper motion parameters. Its PanSTARRS images do not show resolved structure at this scale. A follow-up observation with spatially-resolved spectroscopy is necessary to determine whether this object is a dual or offset AGN. 

We have also checked the 84 AGN with a single Gaia match within 3\arcsec\ but the positional difference between Gaia and SDSS is more than 0\farcs3. Their distributions in Gaia color, proper motion and parallax are indistinguishable from the bulk of the singly-matched AGN. Therefore we conclude that most of them should be due to positional uncertainties in SDSS, and the fraction of these objects is consistent with the expectation from statistical uncertainties given the stated 0\farcs1 positional accuracy of the SDSS. 

For our parent AGN sample (29,769 objects) or the clean sample (8,210 objects), there is a small fraction ($\sim 3\%$) of objects with significant ($>3\sigma$) detection of either parallax or proper motion from Gaia DR2. At this point it is unclear if these parallax and proper motion measurements are due to {\tt varstrometry} of real off-nucleus (or dual) AGN, or due to systematics in Gaia DR2 astrometry. Importantly, the parallax or proper motion measurements in these objects are comparable to or smaller than the corresponding {\tt astrometric\_excess\_noise}, therefore they do not affect our statistical results in \S\ref{sec:gaia}.

\section{Conclusions}\label{sec:con}

Using the {\tt varstrometry} technique \citep[Paper I, also see][]{Wielen_1996,Shen2012a,Liuyuan2015} and Gaia DR2 data, we have constrained the population of sub-kpc-scale, off-nucleus AGN among low-redshift ($0.3<z<0.8$) AGN. We find that low-redshift AGN are well-centered to $<1$ kpc, $<500$ pc and $<100$ pc for 99\%, 90\% and 40\% of the population, respectively. These constraints can be further tightened if we restrict the AGN sample to the most variable subset, with $99.6\%$ of them being well-centered to below 500 parsec from the galactic center. These are the first ever statistical constraints on the off-nucleus AGN population at any redshift. 

We expect future Gaia releases to further improve these statistical constraints. First and foremost, full time series of photocenter and flux measurements can be used to directly compute the astrometric and photometric RMS variability, and to measure the expected linear correlation between the instantaneous photocenter and flux (see discussions in Paper I). Moreover, improved treatment of extended sources in future Gaia releases, increased temporal baselines and potential reduction in modeling systematics will further enhance the power of using Gaia {\tt astrometric\_excess\_noise} to constrain the upper limit of the off-nucleus distance. 

While we have ruled out a large population of off-nucleus AGN beyond a few hundred parsec in the low-reshift regime, galaxy mergers (and presumably SMBH coalescence) are more frequent at high redshift. It is therefore reasonable to expect a higher fraction of off-nucleus AGN from the inspiraling phase or from recoiling SMBHs at higher redshifts. It is possible to extend our analysis to high-redshift quasars. One caveat is that the host fraction (usually unknown) and photometric variability will be lower in high-redshift and high-luminosity quasars than the AGN studied here, leading to less stringent upper limits (cf. Eqn.\ \ref{eqn:key1}). Nevertheless, this will still be an important application of {\tt varstrometry} and upcoming Gaia data releases as a complementary method to systematically search for and constrain the off-nucleus SMBH population. On the other hand, a recent study by \citet{Reines_etal_2019} reported a large fraction of off-nucleus radio-detected AGN in nearby dwarf galaxies. These dwarf galaxies are much less massive and have much shallower galactic potentials than our AGN, which may qualitatively explain the different results in the two studies.

The constraints on sub-kpc off-nucleus AGN from Gaia with {\tt varstrometry} can be compared with predictions from cosmological simulations of inspiraling AGN pairs \citep[e.g.,][]{Tremmel2018} and recoiling AGN \citep[e.g.,][]{Blecha2016}, where the observed AGN luminosity function is reproduced as a basic validation of the simulations. Such comparisons in turn will test the assumptions in the simulations/post-processing, such as AGN fueling recipes, distribution of recoil velocities, and correlations between inspiraling/recoiling SMBHs and host galaxy properties. These comparisons can be performed by matching the simulation/post-processing outputs with the observational parameters, such as magnitude and redshift ranges, host-AGN contrast, host galaxy types, and the spatial scales that Gaia {\tt varstrometry} is sensitive to. The comparison between theoretical predictions and observational constraints on the off-nucleus AGN population is still in its infancy, and we plan to conduct a dedicated comparison using Gaia {\tt varstrometry} constraints in the sub-kpc regime in future work. 


\acknowledgments

We thank the anonymous referee for constructive comments, and Scott Tremaine for useful discussion. We gratefully acknowledge the Heising-Simons Foundation and Research Corporation for Science Advancement for their support to this work. YS acknowledges partial support from an Alfred P. Sloan Research Fellowship and NSF grant AST-1715579. XL acknowledges support from the University of Illinois Campus Research Board.

\bibliography{ref}

\begin{thebibliography}{}
\expandafter\ifx\csname natexlab\endcsname\relax\def\natexlab#1{#1}\fi
\providecommand{\url}[1]{\href{#1}{#1}}

\bibitem[{{Baker} {et~al.}(2006){Baker}, {Centrella}, {Choi}, {Koppitz}, {van
  Meter}, \& {Miller}}]{Baker2006}
{Baker}, J.~G., {Centrella}, J., {Choi}, D.-I., {et~al.} 2006, \apjl, 653, L93

\bibitem[{{Barrows} {et~al.}(2016){Barrows}, {Comerford}, {Greene}, \&
  {Pooley}}]{Barrows_etal_2016}
{Barrows}, R.~S., {Comerford}, J.~M., {Greene}, J.~E., \& {Pooley}, D. 2016,
  \apj, 829, 37

\bibitem[{{Barth} {et~al.}(2008){Barth}, {Bentz}, {Greene}, \& {Ho}}]{barth08}
{Barth}, A.~J., {Bentz}, M.~C., {Greene}, J.~E., \& {Ho}, L.~C. 2008, \apjl,
  683, L119

\bibitem[{{Blecha} {et~al.}(2011){Blecha}, {Cox}, {Loeb}, \&
  {Hernquist}}]{Blecha2011}
{Blecha}, L., {Cox}, T.~J., {Loeb}, A., \& {Hernquist}, L. 2011, \mnras, 412,
  2154

\bibitem[{{Blecha} {et~al.}(2016){Blecha}, {Sijacki}, {Kelley}, {Torrey},
  {Vogelsberger}, {Nelson}, {Springel}, {Snyder}, \& {Hernquist}}]{Blecha2016}
{Blecha}, L., {Sijacki}, D., {Kelley}, L.~Z., {et~al.} 2016, \mnras, 456, 961

\bibitem[{{Bogdanovi{\'c}} {et~al.}(2007){Bogdanovi{\'c}}, {Reynolds}, \&
  {Miller}}]{bogdanovic07}
{Bogdanovi{\'c}}, T., {Reynolds}, C.~S., \& {Miller}, M.~C. 2007, \apjl, 661,
  L147

\bibitem[{{Campanelli} {et~al.}(2007){Campanelli}, {Lousto}, {Zlochower}, \&
  {Merritt}}]{campanelli07}
{Campanelli}, M., {Lousto}, C.~O., {Zlochower}, Y., \& {Merritt}, D. 2007,
  Physical Review Letters, 98, 231102

\bibitem[{{Chambers} {et~al.}(2016){Chambers}, {Magnier}, {Metcalfe},
  {Flewelling}, {Huber}, {Waters}, {Denneau}, {Draper}, {Farrow}, {Finkbeiner},
  {Holmberg}, {Koppenhoefer}, {Price}, {Saglia}, {Schlafly}, {Smartt},
  {Sweeney}, {Wainscoat}, {Burgett}, {Grav}, {Heasley}, {Hodapp}, {Jedicke},
  {Kaiser}, {Kudritzki}, {Luppino}, {Lupton}, {Monet}, {Morgan}, {Onaka},
  {Stubbs}, {Tonry}, {Banados}, {Bell}, {Bender}, {Bernard}, {Botticella},
  {Casertano}, {Chastel}, {Chen}, {Chen}, {Cole}, {Deacon}, {Frenk},
  {Fitzsimmons}, {Gezari}, {Goessl}, {Goggia}, {Goldman}, {Grebel}, {Hambly},
  {Hasinger}, {Heavens}, {Heckman}, {Henderson}, {Henning}, {Holman}, {Hopp},
  {Ip}, {Isani}, {Keyes}, {Koekemoer}, {Kotak}, {Long}, {Lucey}, {Liu},
  {Martin}, {McLean}, {Morganson}, {Murphy}, {Nieto-Santisteban}, {Norberg},
  {Peacock}, {Pier}, {Postman}, {Primak}, {Rae}, {Rest}, {Riess}, {Riffeser},
  {Rix}, {Roser}, {Schilbach}, {Schultz}, {Scolnic}, {Szalay}, {Seitz},
  {Shiao}, {Small}, {Smith}, {Soderblom}, {Taylor}, {Thakar}, {Thiel},
  {Thilker}, {Urata}, {Valenti}, {Walter}, {Watters}, {Werner}, {White},
  {Wood-Vasey}, \& {Wyse}}]{Chambers2016}
{Chambers}, K.~C., {Magnier}, E.~A., {Metcalfe}, N., {et~al.} 2016, arXiv
  e-prints, arXiv:1612.05560

\bibitem[{{Civano} {et~al.}(2010){Civano}, {Elvis}, {Lanzuisi}, {Jahnke},
  {Zamorani}, {Blecha}, {Bongiorno}, {Brusa}, {Comastri}, {Hao}, {Leauthaud},
  {Loeb}, {Mainieri}, {Piconcelli}, {Salvato}, {Scoville}, {Trump}, {Vignali},
  {Aldcroft}, {Bolzonella}, {Bressert}, {Finoguenov}, {Fruscione}, {Koekemoer},
  {Cappelluti}, {Fiore}, {Giodini}, {Gilli}, {Impey}, {Lilly}, {Lusso},
  {Puccetti}, {Silverman}, {Aussel}, {Capak}, {Frayer}, {Le Floch},
  {McCracken}, {Sanders}, {Schiminovich}, \& {Taniguchi}}]{Civano_etal_2010}
{Civano}, F., {Elvis}, M., {Lanzuisi}, G., {et~al.} 2010, \apj, 717, 209

\bibitem[{{Comerford} \& {Greene}(2014)}]{Comerford2014}
{Comerford}, J.~M., \& {Greene}, J.~E. 2014, \apj, 789, 112

\bibitem[{{Comerford} {et~al.}(2009){Comerford}, {Griffith}, {Gerke}, {Cooper},
  {Newman}, {Davis}, \& {Stern}}]{Comerford_etal_2009b}
{Comerford}, J.~M., {Griffith}, R.~L., {Gerke}, B.~F., {et~al.} 2009, \apjl,
  702, L82

\bibitem[{{Comerford} {et~al.}(2015){Comerford}, {Pooley}, {Barrows}, {Greene},
  {Zakamska}, {Madejski}, \& {Cooper}}]{Comerford2015}
{Comerford}, J.~M., {Pooley}, D., {Barrows}, R.~S., {et~al.} 2015, \apj, 806,
  219

\bibitem[{{Gaia Collaboration} {et~al.}(2016){Gaia Collaboration}, {Prusti},
  {de Bruijne}, {Brown}, {Vallenari}, {Babusiaux}, {Bailer-Jones}, {Bastian},
  {Biermann}, {Evans}, {Eyer}, {Jansen}, {Jordi}, {Klioner}, {Lammers},
  {Lindegren}, {Luri}, {Mignard}, {Milligan}, {Panem}, {Poinsignon},
  {Pourbaix}, {Randich}, {Sarri}, {Sartoretti}, {Siddiqui}, {Soubiran},
  {Valette}, {van Leeuwen}, {Walton}, {Aerts}, {Arenou}, {Cropper}, {Drimmel},
  {H{\o}g}, {Katz}, {Lattanzi}, {O'Mullane}, {Grebel}, {Holland}, {Huc},
  {Passot}, {Bramante}, {Cacciari}, {Casta{\~n}eda}, {Chaoul}, {Cheek}, {De
  Angeli}, {Fabricius}, {Guerra}, {Hern{\'a}ndez}, {Jean-Antoine-Piccolo},
  {Masana}, {Messineo}, {Mowlavi}, {Nienartowicz}, {Ord{\'o}{\~n}ez- Blanco},
  {Panuzzo}, {Portell}, {Richards}, {Riello}, {Seabroke}, {Tanga},
  {Th{\'e}venin}, {Torra}, {Els}, {Gracia- Abril}, {Comoretto},
  {Garcia-Reinaldos}, {Lock}, {Mercier}, {Altmann}, {Andrae}, {Astraatmadja},
  {Bellas-Velidis}, {Benson}, {Berthier}, {Blomme}, {Busso}, {Carry},
  {Cellino}, {Clementini}, {Cowell}, {Creevey}, {Cuypers}, {Davidson}, {De
  Ridder}, {de Torres}, {Delchambre}, {Dell'Oro}, {Ducourant}, {Fr{\'e}mat},
  {Garc{\'\i}a-Torres}, {Gosset}, {Halbwachs}, {Hambly}, {Harrison}, {Hauser},
  {Hestroffer}, {Hodgkin}, {Huckle}, {Hutton}, {Jasniewicz}, {Jordan},
  {Kontizas}, {Korn}, {Lanzafame}, {Manteiga}, {Moitinho}, {Muinonen},
  {Osinde}, {Pancino}, {Pauwels}, {Petit}, {Recio-Blanco}, {Robin}, {Sarro},
  {Siopis}, {Smith}, {Smith}, {Sozzetti}, {Thuillot}, {van Reeven}, {Viala},
  {Abbas}, {Abreu Aramburu}, {Accart}, {Aguado}, {Allan}, {Allasia},
  {Altavilla}, {{\'A}lvarez}, {Alves}, {Anderson}, {Andrei}, {Anglada Varela},
  {Antiche}, {Antoja}, {Ant{\'o}n}, {Arcay}, {Atzei}, {Ayache}, {Bach},
  {Baker}, {Balaguer-N{\'u}{\~n}ez}, {Barache}, {Barata}, {Barbier}, {Barblan},
  {Baroni}, {Barrado y Navascu{\'e}s}, {Barros}, {Barstow}, {Becciani},
  {Bellazzini}, {Bellei}, {Bello Garc{\'\i}a}, {Belokurov}, {Bendjoya},
  {Berihuete}, {Bianchi}, {Bienaym{\'e}}, {Billebaud}, {Blagorodnova},
  {Blanco-Cuaresma}, {Boch}, {Bombrun}, {Borrachero}, {Bouquillon}, {Bourda},
  {Bouy}, {Bragaglia}, {Breddels}, {Brouillet}, {Br{\"u}semeister},
  {Bucciarelli}, {Budnik}, {Burgess}, {Burgon}, {Burlacu}, {Busonero}, {Buzzi},
  {Caffau}, {Cambras}, {Campbell}, {Cancelliere}, {Cantat-Gaudin}, {Carlucci},
  {Carrasco}, {Castellani}, {Charlot}, {Charnas}, {Charvet}, {Chassat},
  {Chiavassa}, {Clotet}, {Cocozza}, {Collins}, {Collins}, {Costigan}, {Crifo},
  {Cross}, {Crosta}, {Crowley}, {Dafonte}, {Damerdji}, {Dapergolas}, {David},
  {David}, {De Cat}, {de Felice}, {de Laverny}, {De Luise}, {De March}, {de
  Martino}, {de Souza}, {Debosscher}, {del Pozo}, {Delbo}, {Delgado},
  {Delgado}, {di Marco}, {Di Matteo}, {Diakite}, {Distefano}, {Dolding}, {Dos
  Anjos}, {Drazinos}, {Dur{\'a}n}, {Dzigan}, {Ecale}, {Edvardsson}, {Enke},
  {Erdmann}, {Escolar}, {Espina}, {Evans}, {Eynard Bontemps}, {Fabre},
  {Fabrizio}, {Faigler}, {Falc{\~a}o}, {Farr{\`a}s Casas}, {Faye}, {Federici},
  {Fedorets}, {Fern{\'a}ndez-Hern{\'a}ndez}, {Fernique}, {Fienga}, {Figueras},
  {Filippi}, {Findeisen}, {Fonti}, {Fouesneau}, {Fraile}, {Fraser}, {Fuchs},
  {Furnell}, {Gai}, {Galleti}, {Galluccio}, {Garabato}, {Garc{\'\i}a-Sedano},
  {Gar{\'e}}, {Garofalo}, {Garralda}, {Gavras}, {Gerssen}, {Geyer}, {Gilmore},
  {Girona}, {Giuffrida}, {Gomes}, {Gonz{\'a}lez-Marcos},
  {Gonz{\'a}lez-N{\'u}{\~n}ez}, {Gonz{\'a}lez-Vidal}, {Granvik}, {Guerrier},
  {Guillout}, {Guiraud}, {G{\'u}rpide}, {Guti{\'e}rrez-S{\'a}nchez}, {Guy},
  {Haigron}, {Hatzidimitriou}, {Haywood}, {Heiter}, {Helmi}, {Hobbs},
  {Hofmann}, {Holl}, {Holland}, {Hunt}, {Hypki}, {Icardi}, {Irwin}, {Jevardat
  de Fombelle}, {Jofr{\'e}}, {Jonker}, {Jorissen}, {Julbe}, {Karampelas},
  {Kochoska}, {Kohley}, {Kolenberg}, {Kontizas}, {Koposov}, {Kordopatis},
  {Koubsky}, {Kowalczyk}, {Krone-Martins}, {Kudryashova}, {Kull}, {Bachchan},
  {Lacoste-Seris}, {Lanza}, {Lavigne}, {Le Poncin-Lafitte}, {Lebreton},
  {Lebzelter}, {Leccia}, {Leclerc}, {Lecoeur-Taibi}, {Lemaitre}, {Lenhardt},
  {Leroux}, {Liao}, {Licata}, {Lindstr{\o}m}, {Lister}, {Livanou}, {Lobel},
  {L{\"o}ffler}, {L{\'o}pez}, {Lopez-Lozano}, {Lorenz}, {Loureiro},
  {MacDonald}, {Magalh{\~a}es Fernandes}, {Managau}, {Mann}, {Mantelet},
  {Marchal}, {Marchant}, {Marconi}, {Marie}, {Marinoni}, {Marrese},
  {Marschalk{\'o}}, {Marshall}, {Mart{\'\i}n-Fleitas}, {Martino}, {Mary},
  {Matijevi{\v{c}}}, {Mazeh}, {McMillan}, {Messina}, {Mestre}, {Michalik},
  {Millar}, {Miranda}, {Molina}, {Molinaro}, {Molinaro}, {Moln{\'a}r},
  {Moniez}, {Montegriffo}, {Monteiro}, {Mor}, {Mora}, {Morbidelli}, {Morel},
  {Morgenthaler}, {Morley}, {Morris}, {Mulone}, {Muraveva}, {Musella},
  {Narbonne}, {Nelemans}, {Nicastro}, {Noval}, {Ord{\'e}novic},
  {Ordieres-Mer{\'e}}, {Osborne}, {Pagani}, {Pagano}, {Pailler}, {Palacin},
  {Palaversa}, {Parsons}, {Paulsen}, {Pecoraro}, {Pedrosa}, {Pentik{\"a}inen},
  {Pereira}, {Pichon}, {Piersimoni}, {Pineau}, {Plachy}, {Plum}, {Poujoulet},
  {Pr{\v{s}}a}, {Pulone}, {Ragaini}, {Rago}, {Rambaux}, {Ramos-Lerate},
  {Ranalli}, {Rauw}, {Read}, {Regibo}, {Renk}, {Reyl{\'e}}, {Ribeiro},
  {Rimoldini}, {Ripepi}, {Riva}, {Rixon}, {Roelens}, {Romero-G{\'o}mez},
  {Rowell}, {Royer}, {Rudolph}, {Ruiz-Dern}, {Sadowski}, {Sagrist{\`a}
  Sell{\'e}s}, {Sahlmann}, {Salgado}, {Salguero}, {Sarasso}, {Savietto},
  {Schnorhk}, {Schultheis}, {Sciacca}, {Segol}, {Segovia}, {Segransan},
  {Serpell}, {Shih}, {Smareglia}, {Smart}, {Smith}, {Solano}, {Solitro},
  {Sordo}, {Soria Nieto}, {Souchay}, {Spagna}, {Spoto}, {Stampa}, {Steele},
  {Steidelm{\"u}ller}, {Stephenson}, {Stoev}, {Suess}, {S{\"u}veges}, {Surdej},
  {Szabados}, {Szegedi-Elek}, {Tapiador}, {Taris}, {Tauran}, {Taylor},
  {Teixeira}, {Terrett}, {Tingley}, {Trager}, {Turon}, {Ulla}, {Utrilla},
  {Valentini}, {van Elteren}, {Van Hemelryck}, {van Leeuwen}, {Varadi},
  {Vecchiato}, {Veljanoski}, {Via}, {Vicente}, {Vogt}, {Voss}, {Votruba},
  {Voutsinas}, {Walmsley}, {Weiler}, {Weingrill}, {Werner}, {Wevers},
  {Whitehead}, {Wyrzykowski}, {Yoldas}, {{\v{Z}}erjal}, {Zucker}, {Zurbach},
  {Zwitter}, {Alecu}, {Allen}, {Allende Prieto}, {Amorim},
  {Anglada-Escud{\'e}}, {Arsenijevic}, {Azaz}, {Balm}, {Beck}, {Bernstein},
  {Bigot}, {Bijaoui}, {Blasco}, {Bonfigli}, {Bono}, {Boudreault}, {Bressan},
  {Brown}, {Brunet}, {Bunclark}, {Buonanno}, {Butkevich}, {Carret}, {Carrion},
  {Chemin}, {Ch{\'e}reau}, {Corcione}, {Darmigny}, {de Boer}, {de Teodoro}, {de
  Zeeuw}, {Delle Luche}, {Domingues}, {Dubath}, {Fodor}, {Fr{\'e}zouls},
  {Fries}, {Fustes}, {Fyfe}, {Gallardo}, {Gallegos}, {Gardiol}, {Gebran},
  {Gomboc}, {G{\'o}mez}, {Grux}, {Gueguen}, {Heyrovsky}, {Hoar}, {Iannicola},
  {Isasi Parache}, {Janotto}, {Joliet}, {Jonckheere}, {Keil}, {Kim},
  {Klagyivik}, {Klar}, {Knude}, {Kochukhov}, {Kolka}, {Kos}, {Kutka}, {Lainey},
  {LeBouquin}, {Liu}, {Loreggia}, {Makarov}, {Marseille}, {Martayan},
  {Martinez-Rubi}, {Massart}, {Meynadier}, {Mignot}, {Munari}, {Nguyen},
  {Nordlander}, {Ocvirk}, {O'Flaherty}, {Olias Sanz}, {Ortiz}, {Osorio},
  {Oszkiewicz}, {Ouzounis}, {Palmer}, {Park}, {Pasquato}, {Peltzer}, {Peralta},
  {P{\'e}turaud}, {Pieniluoma}, {Pigozzi}, {Poels}, {Prat}, {Prod'homme},
  {Raison}, {Rebordao}, {Risquez}, {Rocca-Volmerange}, {Rosen}, {Ruiz-Fuertes},
  {Russo}, {Sembay}, {Serraller Vizcaino}, {Short}, {Siebert}, {Silva},
  {Sinachopoulos}, {Slezak}, {Soffel}, {Sosnowska}, {Strai{\v{z}}ys}, {ter
  Linden}, {Terrell}, {Theil}, {Tiede}, {Troisi}, {Tsalmantza}, {Tur},
  {Vaccari}, {Vachier}, {Valles}, {Van Hamme}, {Veltz}, {Virtanen}, {Wallut},
  {Wichmann}, {Wilkinson}, {Ziaeepour}, \& {Zschocke}}]{Gaia2016}
{Gaia Collaboration}, {Prusti}, T., {de Bruijne}, J.~H.~J., {et~al.} 2016,
  \aap, 595

\bibitem[{{Hwang} {et~al.}(2019){Hwang}, {Shen}, {Zakamska}, \&
  {Liu}}]{Hwang_etal_2019}
{Hwang}, H.-C., {Shen}, Y., {Zakamska}, N., \& {Liu}, X. 2019, arXiv e-prints,
  arXiv:1908.02292

\bibitem[{{Kim} {et~al.}(2017){Kim}, {Yoon}, {Privon}, {Evans}, {Harvey},
  {Stierwalt}, \& {Kim}}]{Kim_etal_2017}
{Kim}, D.~C., {Yoon}, I., {Privon}, G.~C., {et~al.} 2017, \apj, 840, 71

\bibitem[{{Komossa}(2012)}]{Komossa2012}
{Komossa}, S. 2012, Advances in Astronomy, 2012, 364973

\bibitem[{{Komossa} \& {Merritt}(2008)}]{Komossa2008}
{Komossa}, S., \& {Merritt}, D. 2008, \apjl, 689, L89

\bibitem[{{Lindegren} {et~al.}(2012){Lindegren}, {Lammers}, {Hobbs},
  {O'Mullane}, {Bastian}, \& {Hern{\'a}ndez}}]{Lindegren_etal_2012}
{Lindegren}, L., {Lammers}, U., {Hobbs}, D., {et~al.} 2012, \aap, 538, A78

\bibitem[{{Liu} {et~al.}(2014){Liu}, {Shen}, {Bian}, {Loeb}, \&
  {Tremaine}}]{Liu2014}
{Liu}, X., {Shen}, Y., {Bian}, F., {Loeb}, A., \& {Tremaine}, S. 2014, \apj,
  789, 140

\bibitem[{{Liu} {et~al.}(2011){Liu}, {Shen}, {Strauss}, \&
  {Hao}}]{Liu_etal_2011}
{Liu}, X., {Shen}, Y., {Strauss}, M.~A., \& {Hao}, L. 2011, \apj, 737, 101

\bibitem[{{Liu}(2015)}]{Liuyuan2015}
{Liu}, Y. 2015, \aap, 580, A133

\bibitem[{{Loeb}(2007)}]{loeb07}
{Loeb}, A. 2007, Physical Review Letters, 99, 041103

\bibitem[{{Merritt} {et~al.}(2004){Merritt}, {Milosavljevi{\'c}}, {Favata},
  {Hughes}, \& {Holz}}]{merritt04}
{Merritt}, D., {Milosavljevi{\'c}}, M., {Favata}, M., {Hughes}, S.~A., \&
  {Holz}, D.~E. 2004, \apjl, 607, L9

\bibitem[{{Miller} \& {Krolik}(2013)}]{Miller2013}
{Miller}, M.~C., \& {Krolik}, J.~H. 2013, \apj, 774, 43

\bibitem[{{Petrov} {et~al.}(2019){Petrov}, {Kovalev}, \&
  {Plavin}}]{Petrov_etal_2019}
{Petrov}, L., {Kovalev}, Y.~Y., \& {Plavin}, A.~V. 2019, \mnras, 482, 3023

\bibitem[{{Pourbaix} {et~al.}(2003){Pourbaix}, {Platais}, {Detournay},
  {Jorissen}, {Knapp}, \& {Makarov}}]{Pourbaix_etal_2003}
{Pourbaix}, D., {Platais}, I., {Detournay}, S., {et~al.} 2003, \aap, 399, 1167

\bibitem[{{Raffai} {et~al.}(2016){Raffai}, {Haiman}, \& {Frei}}]{Raffai2016}
{Raffai}, P., {Haiman}, Z., \& {Frei}, Z. 2016, \mnras, 455, 484

\bibitem[{{Reines} {et~al.}(2019){Reines}, {Condon}, {Darling}, \&
  {Greene}}]{Reines_etal_2019}
{Reines}, A., {Condon}, J., {Darling}, J., \& {Greene}, J. 2019, arXiv
  e-prints, arXiv:1909.04670

\bibitem[{{Reyes} {et~al.}(2008){Reyes}, {Zakamska}, {Strauss}, {Green},
  {Krolik}, {Shen}, {Richards}, {Anderson}, \& {Schneider}}]{Reyes_etal_2008}
{Reyes}, R., {Zakamska}, N.~L., {Strauss}, M.~A., {et~al.} 2008, \aj, 136, 2373

\bibitem[{{Runnoe} {et~al.}(2017){Runnoe}, {Eracleous}, {Pennell}, {Mathes},
  {Boroson}, {Sigur{\dh}sson}, {Bogdanovi{\'c}}, {Halpern}, {Liu}, \&
  {Brown}}]{Runnoe_etal_2017}
{Runnoe}, J.~C., {Eracleous}, M., {Pennell}, A., {et~al.} 2017, \mnras, 468,
  1683

\bibitem[{{Shen}(2012)}]{Shen2012a}
{Shen}, Y. 2012, \apj, 757, 152

\bibitem[{{Sun} \& {Shen}(2015)}]{Sun_Shen_2015}
{Sun}, J., \& {Shen}, Y. 2015, \apjl, 804, L15

\bibitem[{{Tremmel} {et~al.}(2018){Tremmel}, {Governato}, {Volonteri}, {Quinn},
  \& {Pontzen}}]{Tremmel2018}
{Tremmel}, M., {Governato}, F., {Volonteri}, M., {Quinn}, T.~R., \& {Pontzen},
  A. 2018, \mnras, 475, 4967

\bibitem[{{Vanden Berk} {et~al.}(2006){Vanden Berk}, {Shen}, {Yip},
  {Schneider}, {Connolly}, {Burton}, {Jester}, {Hall}, {Szalay}, \&
  {Brinkmann}}]{Vandenberk_etal_2006}
{Vanden Berk}, D.~E., {Shen}, J., {Yip}, C.-W., {et~al.} 2006, \aj, 131, 84

\bibitem[{{Wielen}(1996)}]{Wielen_1996}
{Wielen}, R. 1996, \aap, 314, 679

\end{thebibliography}

\end{document}